\documentclass[prb,aps,twocolumn]{revtex4} %showpacs,
\usepackage{amsmath}
\usepackage{graphicx}
\usepackage{dcolumn}

\begin{document}

\title{Superconducting films with antidot arrays - novel behavior of the critical current}
\author{G. R. Berdiyorov}
\email{golibjon.berdiyorov@ua.ac.be}
\author{M. V. Milo\v{s}evi\'{c}}
\author{F. M. Peeters}
\email{francois.peeters@ua.ac.be}

\affiliation{Departement Fysica, Universiteit Antwerpen,
Groenenborgerlaan 171, B-2020 Antwerpen, Belgium}

\date{\today}

\begin{abstract}

Novel behavior of the critical current density $j_{c}$ of a
regularly perforated superconducting film is found, as a function
of applied magnetic field $H$. Previously pronounced peaks of
$j_{c}$ at matching fields were always found to decrease with
increasing $H$. Here we found a {\it reversal of this behavior}
for particular geometrical parameters of the antidot lattice
and/or temperature. This new phenomenon is due to a strong
``caging'' of interstitial vortices between the pinned ones. We
show that this vortex-vortex interaction can be further tailored
by an appropriate choice of the superconducting material,
described by the Ginzburg-Landau parameter $\kappa$. In effective
type-I samples we predict that the peaks in $j_{c}(H)$ at the
matching fields are transformed into a {\it step-like behavior}.

\end{abstract}


\maketitle

\section{Introduction}

For practical applications of superconducting (SC) materials, the
increase and, more generally, control of the critical current in
SC samples are of great importance. In recent years much attention
was given to the investigation of superconducting films patterned
with a regular array of microholes (antidots), which have a
profound influence on both the critical current and the critical
magnetic field \cite{baert,mosh1,mosh2,mosh3,mosh4}. Due to the
collective pinning to the regular antidot array, vortices are
forced to form rigid lattices when their number ``matches''
integer and fractional multiples of the number of pinning sites at
fields $H_{n}=n\Phi_{0}/S$, and $H_{p/q}=\frac{p}{q}\Phi_{0}/S$
(where $n,p,q$ are integers) respectively, where $\Phi_{0}$ is the
flux quantum, and $S$ is the area of the unit cell of the antidot
lattice. This locking between the pinning array and the vortex
lattice is responsible for the reduced mobility of the vortices in
applied drive and consequently the increased critical current at
integer and fractional matching fields, was confirmed both by
experiments (imaging \cite{harada}, magnetization and transport
measurements \cite{baert,mosh1}) and molecular dynamics
simulations \cite{Rei}.

However, regardless on the imposed pinning profile, the vortices
at interstitial sites always have high mobility
\cite{mosh4,khalfin}, show different dynamic regimes from the
pinned ones, and their appearance is followed by a sharp drop in
the critical current \cite{baert}. In this respect, the maximal
occupation number of the antidots (saturation number $n_{s}$)
becomes very important for any study of the critical current. In
an early theoretical work, Mkrtchyan and Schmidt \cite{mkrt} have
shown that $n_{s}$ depends only on the size of the holes. However,
in the case of periodic pinning arrays this number depends also on
the proximity of the holes, on temperature and on the applied
field \cite{baert,mosh4,buzdin,doria1}.

Besides the pinning strength of the artificial lattice,
vortex-vortex interactions are crucial for vortex dynamics. Most
of the experiments on perforated superconducting films are carried
out in the effective type-II regime
($\kappa_{\ast}=2\kappa^2\big/d\gg 1/\sqrt{2}$, with $\kappa$
being the Ginzburg-Landau (GL) parameter, and $d$ being the
thickness of the SC film scaled to the coherence length $\xi$),
where vortices act like charged point-particles, and their
interaction with the periodic pinning potential can be described
using molecular dynamics simulations (MDS) \cite{Rei}. However,
the overlap of vortex cores (with sizes $\sim \xi$), and the exact
shape of the inter-vortex interaction (depending on the material
properties reflected through $\kappa$), which are neglected in
MDS, may significantly modify the equilibrium vortex structures
and consequently the critical current.

In the present letter we investigate the critical current $j_c$ of
superconducting films with regular arrays of square antidots,
taking into account all parameters relevant to the SC state,
within the non-linear Ginzburg-Landau theory. This formalism
allows us to analyze the $j_c$ dependence on the geometrical
parameters of the sample, material (even type-I) and temperature,
and compare our results with existing experiments.

\section{Theoretical approach}

\begin{figure*} \centering
\vspace{0cm}
\includegraphics[scale=0.6]{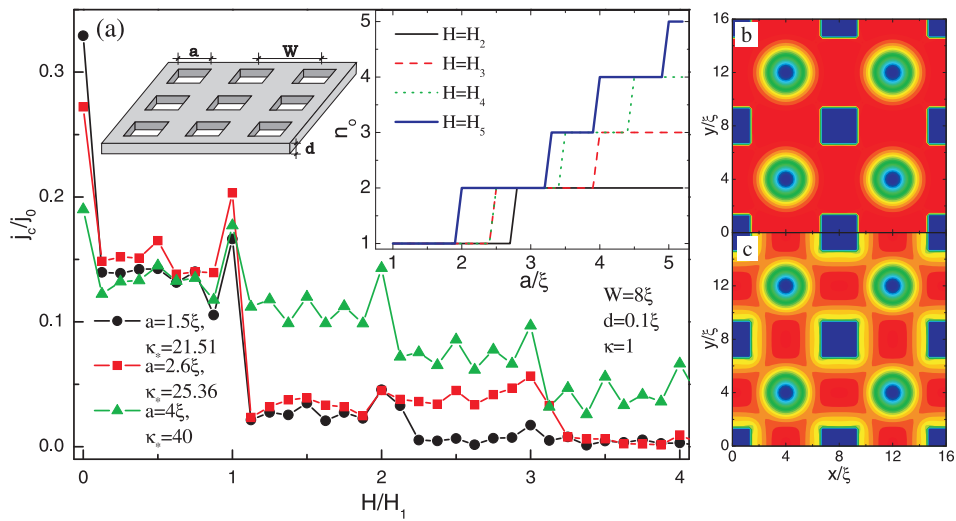}
\vspace{0cm} \caption{(Color online) The critical current density
$j_{c}$ (a) (in units of $j_{0}=cH_{c2}\xi/4\pi \lambda^{2}$) as a
function of the applied magnetic field $H$ (in units of the first
matching field $H_{1}$) for different antidot-sizes and fixed
period $W=8\xi$, and the contour plot of the Cooper-pair density
at $H=H_{2}$ (b) and at $H=H_{3}$ (c) for $R=2.6\xi$. The insets
show the schematic view of the sample and the antidot occupation
number $n_{o}$ as a function of the antidot-size $a$ at different
matching fields (\textit{i.e.} for $2-5$ vortices per unit cell).}
\label{fig1}
\end{figure*}

We consider a thin superconducting film (of thickness $d$) with a
regular lattice of square holes (side $a$, period $W$) immersed in
an insulating media with a perpendicular uniform applied field $H$
(see the inset of Fig. \ref{fig1}). To describe this system, we
solved the nonlinear Ginzburg-Landau (GL) equations for the order
parameter $\psi$ and the vector potential $\vec{A}$ (in
dimensionless units, and with temperature $T$ taken explicitly in
units of the zero field critical temperature $T_{c0}$, see Ref.
\cite{schw} for more details):

\begin{equation}
\left( -i\vec{\nabla}-\vec{A}\right) ^{2}\psi =\psi \left(
1-T-\left| \psi \right| ^{2}\right) , \label{eq1}
\end{equation}
\begin{equation}
-\kappa_{\ast}\Delta\vec{A}=\frac{1}{i}\left( \psi ^{\ast
}\vec{\nabla}\psi -\psi \vec{\nabla }\psi ^{\ast }\right) -\left|
\psi \right| ^{2}\vec{A}.  \label{eq2}
\end{equation}
The magnitude of the applied field $H$ is determined by the number
of flux quanta piercing through the simulation region $W_s\times
W_s$. On the hole-edges we used boundary conditions corresponding
to zero normal component of the superconducting current. Periodic
boundary conditions \cite{doria} are used around the square
simulation region:
$\vec{A}(\vec{r}+\vec{l}_{i})=\vec{A}(\vec{r})+\vec{\nabla}\zeta_{i}(\vec{r})$
and $\psi(\vec{r}+\vec{l}_{i})=\psi \exp(2\pi i\zeta
_{i}(\vec{r})/\Phi _{0})$, where $\vec{l}_{i}$ ($i=x,y$) are
lattice vectors and $\zeta_i$ is the gauge potential. We use the
Landau gauge $\vec{A}_0=Hx\vec{e}_y$ for the external vector
potential and $\zeta_x=HW_sy+C_x$, $\zeta_y=C_y$, with $C_x$,
$C_y$, being constants. Generally speaking, depending on the
geometry of the antidot lattice, one must minimize the energy with
respect to the latter coefficients. The size of the supercell in
our calculation is typically $4\times 4$ unit cells (containing 16
holes, \textit{i.e.} $W_s=4W$). We solved the system of Eqs.
(\ref{eq1})-(\ref{eq2}) self-consistently using the numerical
technique of Ref. \cite{schw}. In order to calculate the critical
current, first we determine the ground vortex-state for a given
applied magnetic field after multiple starts from a randomly
generated initial Cooper-pair distribution. Then the applied
current in the $x$-direction is simulated by adding a constant
$A_{cx}$ to the vector potential of the applied external field
\cite{misko1}. Note that the current $j_x$ in the sample resulting
from the applied $A_{cx}$ is obtained after integration of the
$x$-component of the induced supercurrents (calculated from Eq.
(\ref{eq2})) over the $y=const.$ cross-section of the sample. With
increasing $A_{cx}$, the critical current is reached when a
stationary solution for Eqs. (\ref{eq1})-(\ref{eq2}) ceases to
exist (\textit{i.e.} vortices are driven in motion by the Lorentz
force).

\section{Influence of geometrical parameters of the sample}

\begin{figure}[b] \centering
\vspace{0cm}
\includegraphics[scale=0.45]{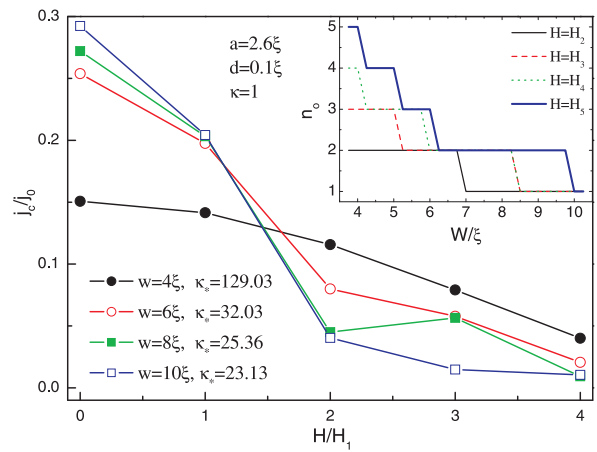}
\vspace{0cm} \caption{(Color online) The critical current density
$j_{c}$ of the superconducting film at matching fields for
different periods of the antidot lattice. The inset shows the
antidot occupation number $n_{o}$ as a function of $W$ at
different matching fields.} \label{fig2}
\end{figure}

The enforced stability of the vortex lattices in periodically
perforated superconducting samples \cite{harada,novel} at integer
and fractional matching fields leads to pronounced peaks in the
critical current \cite{baert,mosh1,mosh2}. However, the exact
shape of these lattices, and consequently their stability when
locked to the pinning arrays, strongly depend on the parameters of
the sample. For example, while small antidots can pin only one
vortex, in larger holes multi-quanta vortices may become
energetically preferable \cite{buzdin}. This reduces the number of
interstitial vortices, whose higher mobility affects strongly the
critical current of the sample. Therefore, in this section we
investigate the critical current of our sample for different sizes
of the antidots $a$ and antidot lattice period $W$.

Fig. \ref{fig1} shows the critical current density $j_{c}$ of the
sample as a function of the applied magnetic field $H$ for three
different values of the antidot size: $a=1.5\xi$, $a=2.6\xi$ and
$a=4\xi$. The lattice period is $W=8\xi$, the film thickness
$d=0.1\xi$ and GL parameter equals $\kappa=1$ (corresponding
roughly to Pb, Nb, or Al films). Note that, following the
suggestion of Wahl \cite{wahl}, we took into account the influence
of the perforation on the effective GL parameter
$\kappa_{\ast}=2\kappa^{2}/d(1-2a^{2}/W^{2})$. As shown in Fig.
\ref{fig1}, in the absence of applied field the samples with
smaller antidots always have larger $j_{c}$ simply due to more
superconducting material, \textit{i.e.} larger screening. However,
for $H\neq 0$, the critical current depends on the vortex
structure in the sample. For small antidot-size $a=1.5\xi$, where
only one vortex can be captured by each hole (see the inset of
Fig. \ref{fig1}), the $j_{c}(H)$ curve shows the expected maxima
at integer matching fields $H_{1}$, $H_{2}$, $H_{3}$ and $H_{4}$
and at some of the fractional matching fields. As observed before
\cite{baert}, the peaks of the critical current at matching fields
decrease with increasing magnetic field. However, a novel
phenomenon is found for $2.5\leq a/\xi \leq 2.8$: {\it the
critical current at the $3^{rd}$ matching field is larger than the
one at the $2^{nd}$ matching field}. Moreover, the critical
currents for $H_{2}<H<H_{3}$ are higher than those obtained for
fields $H_{1}<H<H_{2}$.

The explanation for this counterintuitive feature of $j_c$ lies in
the hole occupation number $n_{o}$, \textit{i.e.} the number of
vortices inside the hole, and, consequently, in the saturation
number $n_{s}$ of the holes ($n_{s}=n_{o}$ for larger fields). For
holes with $2.5\leq a \leq 2.8$, $n_{s}$ is equal to 1
($n_{s}\approx a/4\xi$ \cite{mkrt}). Therefore, at $H=H_1$, all
vortices are captured by the antidots. Analogously, at $H=H_2$,
besides the pinned vortices, one vortex occupies each interstitial
site. If the same analogy is followed further, one expects two
vortices at each pinning site at the $3^{rd}$ matching field.
However, this depends also on the distance between neighboring
holes. Namely, the period $W$ determines the distance between the
interstitial vortices. In other words, for small $W$, this
distance is small, vortex-vortex repulsion is large, and it may
become energetically more favorable for the system to force one
more flux-line into the hole rather than have two vortices at
interstitial sides (see Figs.~\ref{fig1}(b,c)). Therefore, our
results show that the occupation number $n_{o}$ of the holes in a
lattice {\it depends not only on the size of the hole $a$, but
also on the period $W$ and the number of vortices per unit-cell of
the hole-lattice}. For the parameters given above, $n_o=2$ at
$H=H_{3}$, \textit{i.e.} only one vortex is located at each
interstitial site, as for the case of $H=H_2$. Note that this
interstitial vortex interacts repulsively with the pinned ones.
This interaction is roughly twice as large at $H=H_{3}$ than at
$H=H_2$, and the interstitial vortex becomes effectively
``caged''. This ``caging'' effect has been found for $7\xi \leq
W\leq 8.3\xi$ for the radius $R=2.6\xi$. For $W>8.3\xi$, $n_{o}$
and $n_s$ become independent of $W$ and for $W<7\xi$, the larger
suppression of superconductivity around the holes at $H=H_3$
becomes more significant than ``caging'', and $j_c$ reverses again
in favor of $H=H_2$.

With further increase of hole-size $a$, more vortices are captured
by the holes. Consequently, in order to observe the ``caging''
effect, one needs to consider higher magnetic fields,
\textit{i.e.} for $n_o=4$, at least $H>H_5$ is needed. In any
case, this effect can always be realized at a given magnetic field
by an adequate choice of $a$ and $W$. Fig. \ref{fig2} shows the
critical current density of our sample for different periods of
the antidot lattice $W$ at $a=2.6\xi$. For clarity, we plotted
only the critical current at the matching fields. At $H=0$ and
$H=H_{1}$ (when there are no interstitial vortices) the critical
current is an increasing function of $W$, because of the stronger
screening by larger quantities of the superconducting material.
For higher magnetic fields interstitial vortices nucleate more
easily in a sample with larger $W$, which leads to a reduction of
$j_{c}$. The dependence of $n_{o}$ (and thus also saturation
number $n_{s}$) on $W$ is shown in the inset of Fig. \ref{fig2}:
$n_{o}$ decreases from $n_{o}=5$ to $n_{o}=1$ at $H=H_{5}$, when
we increase $W$ from $4\xi$ to $10\xi$.

\section{Temperature dependence of the critical current}

\begin{figure}[b] \centering
\vspace{0cm}
\includegraphics[scale=0.38]{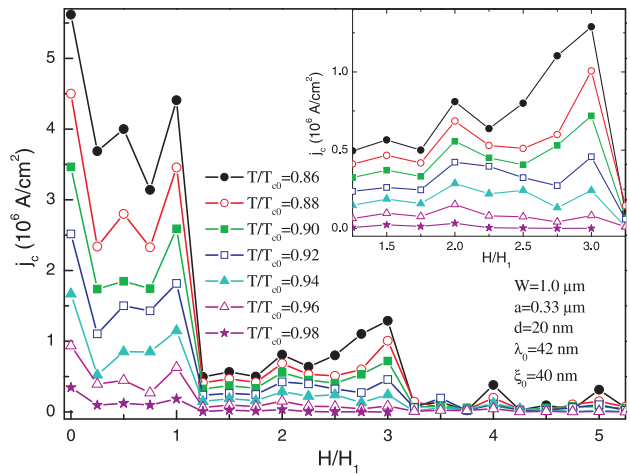}
\vspace{0cm} \caption{(Color online) The temperature dependence of
$j_c(H)$ for a superconducting film containing an antidot array,
at temperatures $T/T_{c0}=0.86-0.98$. The period of the
antidot-lattice is $W=1\mu$m, the antidot size $a=0.33\mu$m, film
thickness $d=20$nm, and $\kappa$ equals $1.05$.} \label{fig3}
\end{figure}

So far, we presented results of our calculations at fixed
temperature, where all units were temperature dependent (e.g.
distances were expressed in $\xi(T)$). In what follows, we
consider a superconducting film with thickness $d=20$nm, interhole
distance $W=1\mu$m, and antidot size $a=0.33\mu$m. We choose the
coherence length $\xi(0)=40$nm and the penetration depth
$\lambda(0)=42$nm, which are typical values for Pb films. We study
the influence of temperature with the help of the right-side term
of Eq. (\ref{eq1}), which actually describes the temperature
dependence of the coherence length
$\xi(T)=\xi(0)/\sqrt{|1-T/T_{c0}|}$ and penetration depth
$\lambda(T)=\lambda(0)/\sqrt{|1-T/T_{c0}|}$.

Fig. \ref{fig3} shows the calculated critical current density of
the sample as a function of the applied field normalized to the
first matching field at temperatures $T/T_{c0}=$0.86, 0.88, 0.9,
0.92, 0.94, 0.96 and 0.98. As discussed in the previous section,
the $j_{c}(H)$ curve shows pronounced maxima at matching fields
with a substantial drop after the number of vortices per unit-cell
exceeds the hole-saturation number (in our case, for $H>H_1$).

As expected, decreasing temperature results in an increase of the
critical current for given magnetic field. However, the
qualitative behavior of the $j_c(H)$ characteristics changes.
Although the $W/a$ ratio remains the same, the size of vortices
and the occupation number of the holes may change with temperature
(as $\xi(T)$ changes). Actually, we found the same vortex
structure at all temperatures $T>0.8T_{c0}$, but $j_c(H_3)$ is
found to be larger than $j_c(H_2)$ only in the temperatures range
$0.8T_{c0}\leq T\leq 0.93 T_{c0}$. At higher temperatures vortices
become larger and suppression of superconductivity around the
holes ``masks'' the critical current enhancement by the ``caging''
effect. This decrease of Cooper-pair density around the holes is
also responsible for the decreased $j_c(H_1)/j_c(0)$ ratio with
increasing temperature. At temperatures lower than $0.8T_{c0}$,
the peaks of $j_c$ again decrease with magnetic field, since
$W/\xi$ significantly increases and its influence on $n_o$
diminishes (see previous section).

Interestingly enough, this $j_c(H_{n+1})$ vs. $j_c(H_{n})$
reversing behavior {\it was recently found experimentally, but not
noticed}. In Fig.~6 of Ref. \cite{mosh3} a clear enhancement of
the critical current at $H/H_{1}=3$ was found which is larger than
the one at $H/H_{2}$ as is similar to the behavior found in
Fig.~\ref{fig3} for $T/T_{c0}<0.94$. A quantitative comparison
between theory and experiment is difficult because of the
different determination of $j_{c}$. In our calculations we assume
normal state as soon as vortices are set in motion, whereas in
transport measurements a certain value of threshold voltage
determines the critical current.

\section{Effective type-I vs. type-II behavior of the critical current}

\begin{figure}[b] \centering
\vspace{0cm}
\includegraphics[scale=0.375]{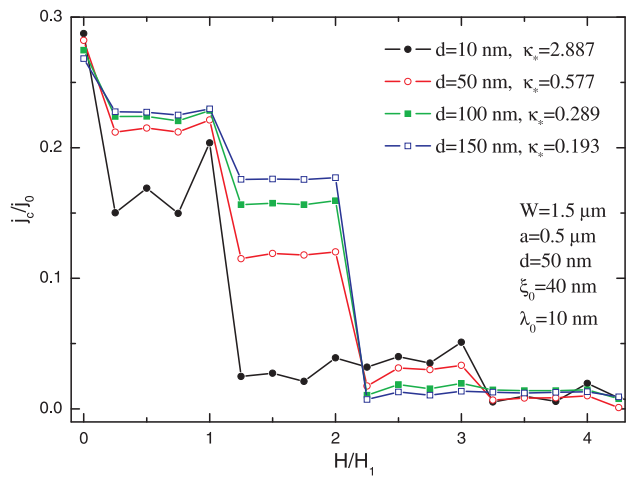}
\vspace{0cm} \caption{(Color online) The critical current density
$j_{c}$ as a function of the applied magnetic field for different
thickness of the sample. Remaining parameters are listed in the
figure.} \label{fig4}
\end{figure}

In the previous sections we have shown that higher mobility of
interstitial vortices leads to a dramatic decrease of the critical
current. Therefore, keeping the interstitial sites of the
superconductor vortex-free is essential for an improvement of
$j_c$. In this respect, we consider effectively type-I
superconductors, where (i) the screening of the magnetic field is
enhanced (\textit{i.e.} vortices will be compressed in the holes),
and (ii) the interaction of vortices becomes attractive (depends
on $\ln{\kappa_{\ast}}$). As an example we considered a sample,
with antidot size $a=0.5\mu$m, lattice period $W=1.5\mu$m,
$\xi_{0}=40$nm and $\lambda_{0}=10$nm. We fine-tuned the effective
vortex-vortex interaction by varying the thickness of the sample,
\textit{i.e.} changing from type-II to type-I behavior with
increasing $d$. Fig. \ref{fig4} shows the critical current of the
sample for four values of the film thickness: $d=10$nm (solid
dots), $d=50$nm (open dots), $d=100$nm (solid squares) and
$d=150$nm (open squares) at $T=0.97T_{c0}$. For $d=10$nm, the
sample is still in the type-II regime ($\kappa_{\ast}=2.887$), and
the critical current shows a peak-like behavior at the matching
fields. The drop in the critical current for $H>H_1$ is caused by
the interstitial vortices which is larger with increasing
$\kappa_{\ast}$. Note that the ``caging'' effect is also present
for these values of the parameters. When the film thickness is
increased ($d=50$nm, $\kappa_{\ast}=0.577$) the critical current
density is higher for $H<H_{2}$. For $H<H_{1}$, this increase is
achieved due to a stronger Meissner effect. At fields
$H_1<H<H_{2}$ the increase of $j_c$ is more apparent, as all
vortices are captured by the holes (except for $d=10$nm, where
$n_{o}=1$). As soon as interstitial vortices appear in the sample
($H>H_2$), $j_{c}$ becomes even smaller than the one for $d=10$nm.
This inversion of $j_{c}$ clearly demonstrates the higher mobility
of interstitial vortices in type-I superconductors.

Looking at the $j_c(H)$ curve as a whole for $d=50$nm, we observed
a pronounced {\it step-like behavior}. Note that in type-I samples
the matching between the number of flux-lines and the number of
antidots does not lead to a peak-like increase of the critical
current. Namely, regardless of their number, additional flux lines
are doubly pinned by the attractive hole-potential and the
attractive interaction with previously pinned vortices. The
``step'' in $j_c(H)$ occurs only when the number of vortices per
hole $n_{o}$ changes (more vortices and consequently a larger
suppression of $\psi$ around the hole), or when interstitial
vortices appear. The effect of an increase of $n_{o}$ diminishes
as $d$ increases (\textit{i.e.} $\kappa_{\ast}$ decreases and
screening increases), or when temperature is lowered (as discussed
before). This tendency is illustrated by solid and open dots in
Fig. \ref{fig4}, and ultimately leads to a {\it two-step $j_c(H)$
curve}, with larger critical current for $H<H_n$, and smaller
$j_c$ for $H>H_n$, where $n=n_o$.

\section{Conclusions}

We studied the critical current of a superconducting film
containing an array of antidots in a uniform magnetic field, as a
function of all relevant parameters. We found that the well-known
$j_{c}$ enhancement by artificial vortex pinning strongly depends
on the antidot occupation number $n_{o}$. The latter is determined
not only by the size of the antidots as commonly believed, but
{\it also by their spacing and the applied field}. As a
consequence, when the parameter conditions are met, the critical
current becomes larger {\it at higher matching fields}, contrary
to the conventional behavior. Such a feature is a result of the
``caging'' of interstitial vortices between the larger number of
pinned ones. This effect is strongly influenced by temperature,
which agrees with a recent experiment.~\cite{mosh3} Additionally,
the interactions in this system can be tailored by the effective
Ginzburg-Landau parameter $\kappa_{\ast}$. In effectively type-I
samples, vortices attract, $j_c$ increases and peaks at matching
fields diminish due to the enforced pinning at all fields. As a
result, a {\it novel step-like behavior} of the critical current
is found, with a sharp drop at higher fields, when interstitial
vortices appear.

\section*{ACKNOWLEDGEMENTS}

This work was supported by the Flemish Science Foundation
(FWO-Vl), the Belgian Science Policy (IUAP), the JSPS-ESF NES
program, and the ESF-AQDJJ network.


\begin{thebibliography}{99}

\bibitem{baert} M. Baert, {\it et al.}, Phys. Rev. Lett.{\bf 74}, 3269 (1995).

\bibitem{mosh1} V.V. Moshchalkov, {\it et al.}, Phys. Rev. B {\bf 57}, 3615 (1998).

\bibitem{mosh2} E. Rosseel, {\it et al.}, Physica C {\bf 282}, 1567 (1997).

\bibitem{mosh3} A.V. Silhanek, {\it et al.}, Phys. Rev. B {\bf 72}, 014507 (2005).

\bibitem{mosh4} A.V. Silhanek, {\it et al.}, Phys. Rev. B {\bf70}, 054515 (2004).

\bibitem{harada} K. Harada, {\it et al.}, Science {\bf 274}, 1167 (1996).

\bibitem{Rei} C. Reichhardt, {\it et al.}, Phys. Rev. B {\bf 54}, 16108 (1997); {\it
ibid.}, Phys. Rev. B {\bf 57}, 7937 (1998); {\it ibid.}, Phys.
Rev. B {\bf 63}, 054510 (2001); {\it ibid.}, Phys. Rev. B {\bf
64}, 014501 (2001); {\it ibid.}, Phys. Rev. B {\bf64}, 052503
(2001).

\bibitem{khalfin} I.B. Khalfin, {\it et al.}, Physica C {\bf 207}, 359 (1993).

\bibitem{mkrt} G.S. Mkrtchyan and V.V. Shmidt, Sov. Phys. JETP {\bf 34}, 195 (1972).

\bibitem{buzdin} A.I. Buzdin, Phys. Rev. B {\bf 47}, 11416 (1993).

\bibitem{doria1} M.M. Doria, {\it et al.}, Physica C {\bf 341}, 1199 (2000);
{\it ibid.}, Phys. Rev. B {\bf 66}, 064519 (2002).

\bibitem{schw} V.A. Schweigert and F.M. Peeters, Phys. Rev. B {\bf 57}, 13817 (1998).

\bibitem{doria} M.M. Doria, {\it et al.}, Phys. Rev. B {\bf 39}, 9573 (1989).

\bibitem{misko1} M.V. Milo\v{s}evi\'{c} and F.M. Peeters, Phys. Rev. Lett. {\bf 93}, 267006 (2004).

\bibitem{novel} G.R. Berdiyorov, {\it et al.}, Phys. Rev. Lett. {\bf 96}, xxx (2006).

\bibitem{wahl} A. Wahl, {\it et al.}, Physica C {\bf 250}, 163 (1995).

\end{thebibliography}
\end{document}